\documentclass[aps,pra,reprint,groupedaddress]{revtex4-1}
\usepackage{graphicx}
\usepackage{color}
\usepackage{bm}
\usepackage{comment}
\usepackage{amssymb, amsmath}

\begin{document}


\title{High-accuracy Ising machine using Kerr-nonlinear parametric oscillators with local four-body interactions}


\author{Taro Kanao}
\email[]{taro.kanao@toshiba.co.jp}
\author{Hayato Goto}
\affiliation{Frontier Research Laboratory, Corporate Research \& Development Center, Toshiba Corporation, 1, Komukai-Toshiba-cho, Saiwai-ku, Kawasaki 212-8582, Japan}


\date{\today}

\begin{abstract}
A two-dimensional array of Kerr-nonlinear parametric oscillators (KPOs) with local four-body interactions is a promising candidate for realizing an Ising machine with all-to-all spin couplings, based on adiabatic quantum computation in the Lechner-Hauke-Zoller (LHZ) scheme. However its performance has been evaluated for only a few KPOs. By numerically simulating more KPOs, here we show that the performance can be dramatically improved by reducing inhomogeneity in photon numbers induced by the four-body interactions. The discrepancies of the photon numbers can be corrected by tuning the detunings of KPOs depending on their positions, without monitoring their states during adiabatic time evolution. This correction can be used independent of the number of KPOs in the LHZ scheme and thus can be applied to large-scale implementation.
\end{abstract}


\maketitle

\section*{Introduction}\label{sec_intro}
Hardware devices for solving the Ising problem~\cite{Barahona1982} have attracted attention because this problem can represent  various combinatorial optimization problems~\cite{Lucas2014}.
Such Ising machines are thus expected to be applied to real-world problems, such as integrated circuit design~\cite{Barahona1988}, computational biology~\cite{Perdomo-Ortiz2012, Li2018}, and financial portfolio management~\cite{Rosenberg2016}.
The Ising machines have been developed in several implementations including quantum annealing~\cite{Kadowaki1998, Das2008} or adiabatic quantum computation~\cite{Farhi2000, Farhi2001, Albash2018} with superconducting circuits~\cite{Johnson2011}, and coherent Ising machines with laser pulses~\cite{Wang2013, Marandi2014, Leleu2017, Yamamoto2017}.
In these devices, quantum effects such as quantum tunneling and quantum fluctuations are expected to be exploited for solving the Ising problem.
Classical Ising machines have also been implemented with digital circuits based on simulated annealing~\cite{Kirkpatrick1983, Yamaoka2016, Aramon2019} and simulated bifurcation~\cite{Goto2019, Tatsumura2019, Zou2020}.

Another adiabatic quantum computation for the Ising problem has been proposed using Kerr-nonlinear parametric oscillators (KPOs)~\cite{Goto2016, Goto2019a}.
A KPO is a parametrically driven oscillator with Kerr nonlinearity, and exhibits a bifurcation~\cite{Dykman2012}.
The bifurcation allows KPOs to represent binary Ising spins.
Furthermore, a KPO can generate non-classical states such as a quantum superposition of coherent states, known as a Schr\"{o}dinger cat state, via a quantum adiabatic bifurcation~\cite{Goto2016, Goto2019a}.
This Ising machine has been referred to as a quantum bifurcation machine (QbM)~\cite{Goto2018a, Goto2019a}.
As driven states in KPOs are used, QbMs can contrast with quantum annealers operated in thermal equilibrium~\cite{Johnson2011, Amin2015}.

Utilizing these features of KPOs, applications other than Ising machines have also been proposed such as gate-based universal quantum computation~\cite{Goto2016a, Puri2017a, Puri2019a}, Boltzmann sampling~\cite{Goto2018a}, studies of quantum phase transitions~\cite{Dykman2018, Rota2019}, and on-demand generation of traveling cat states~\cite{Goto2019b}.
KPOs can be implemented by superconducting circuits like Josephson parametric oscillators~\cite{Yamamoto2008, Lin2014}, and have recently been realized experimentally~\cite{Wang2019, Grimm2019}.

QbMs for the Ising problem with all-to-all spin couplings, which allow us to solve a wider class of real-world problems, have been proposed~\cite{Nigg2017, Puri2017, Zhao2018}.
The architecture in Ref.~\cite{Puri2017} is based on the Lechner-Hauke-Zoller (LHZ) scheme~\cite{Lechner2015, Pastawski2016, Albash2016a, Rocchetto2016, Hartmann2019, Susa2020}, where the all-to-all spin couplings can be realized by a two-dimensional array of KPOs with local four-body interactions.
This architecture is a promising candidate for large-scale implementation of the LHZ scheme, because in the case of QbM, the four-body interaction is realized by four-wave mixing in a single Josephson junction~\cite{Puri2017}.
This simple four-body interaction can be an advantage over quantum annealers in the LHZ scheme with flux qubits~\cite{Chancellor2017} or transmon qubits~\cite{Leib2016}, where a four-body interaction may need complicated circuits with multiple ancilla qubits.

In the previous numerical studies, however, to our knowledge the number of KPOs in the LHZ scheme has been limited up to three~\cite{Note1}, and the performance of its large-scale implementations has not been revealed.
In this paper, we investigate adiabatic quantum computation using a KPO network in the LHZ scheme (LHZ-QbM) with a larger number of KPOs.
By a variational method, we first find that mean photon numbers in the KPOs become inhomogeneous, i.e.~unequal, owing to asymmetry in the LHZ scheme.
This inhomogeneity can degrade accuracy in solving the Ising problem.
We then propose a method to reduce the inhomogeneity by scheduling the detunings of KPOs without monitoring their quantum states during adiabatic time evolution.
Finally we numerically demonstrate that this method can improve the solution accuracy dramatically.
This method is applicable to arbitrary numbers of KPOs in the LHZ-QbM and therefore to its large-scale implementation.

\section*{RESULTS}
\noindent{\bf LHZ-QbM}\\
We first briefly explain the Ising problem and the LHZ scheme, and then introduce an LHZ-QbM.
In the Ising problem~\cite{Barahona1982}, a dimensionless Ising energy, $E_{\rm Ising}=-\!\sum^N_{i=1}\!\sum_{j<i}\!J_{ij}s_is_j$, is minimized with respect to $N$ Ising spins ${\left\{s_i=\pm1\right\}}$, where ${\left\{J_{ij}\right\}}$ represent two-body interactions.

\begin{figure}
	\includegraphics[width=5cm]{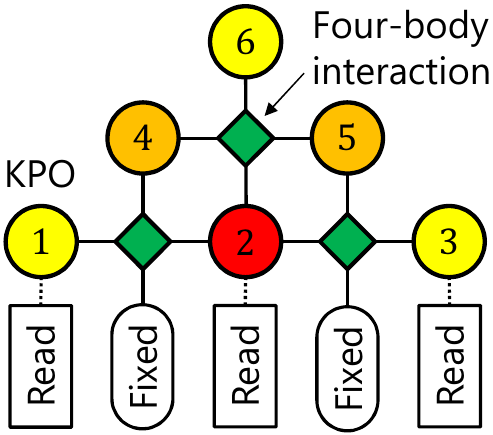}%
	\caption{{\bf A network of Kerr-nonlinear parametric oscillators in the Lechner-Hauke-Zoller  scheme for the four-spin Ising problem.}
		KPOs 1--6 correspond to LHZ spins ${\tilde{s}_k=\pm1}$, while the ancillary LHZ spins (``Fixed'') fixed to 1 are implemented by classical driving.\label{fig_KPONet}}
\end{figure}

In the LHZ scheme~\cite{Lechner2015}, the product of two Ising spins ${s_is_j}$ is mapped to a variable ${\tilde{s}_k=\pm1}$, which we call an LHZ spin.
This mapping reduces the two-body interaction ${J_{ij}}$ to a local external field $J_k$, while increasing the number of the spins to ${L=N(N-1)/2}$.
The additional degrees of freedom are removed by four-body constraints on neighboring LHZ spins:~${\tilde{s}_k\tilde{s}_l\tilde{s}_m\tilde{s}_n=1}$.
In Fig.~\ref{fig_KPONet}, which shows the lattice structure of the LHZ scheme for ${N=4}$ ${(L=6)}$, a four-body constraint is imposed on four LHZ spins connected by a square.
The lower edge of the lattice is terminated by ancillary LHZ spins fixed to $1$.

To satisfy the constraints, terms proportional to ${\tilde{s}_k\tilde{s}_l\tilde{s}_m\tilde{s}_n}$ are added to the Ising energy, resulting in the LHZ energy, $E_{\rm LHZ}=-\!\sum^L_{k=1}\!J_k\tilde{s}_k-C\!\sum_{\langle k, l, m, n\rangle}\!\tilde{s}_k\tilde{s}_l\tilde{s}_m\tilde{s}_n$, to be minimized.
Here the summation in the second term is over all the constraints.
The second term means four-body interactions with their strength $C$.
With sufficiently large $C$, the four-body constraints are satisfied by ${\left\{\tilde{s}_k\right\}}$ minimizing ${E_{\rm LHZ}}$ (see Appendix~\ref{sec_AppA}).

We search for the solution of the Ising problem by embedding the LHZ energy into a network of $L$ KPOs described by a Hamiltonian ${H=H_{\rm KPO}+H_{\rm LHZ}}$,
\begin{eqnarray}
	H_{\rm KPO}&=&\hbar\!\sum^L_{k=1}\!\left[\frac{K}{2}a^{\dagger2}_ka^2_k-\frac{p}{2}\!\left(a^{\dagger2}_k+a^2_k\right)+\Delta_k a^\dagger_ka_k\right],\label{eq_HKPO}\\
	H_{\rm LHZ}&=&-\hbar\xi\!\left(A\!\sum^L_{k=1}\!J_ka^\dagger_k+C\!\sum_{\langle k,l,m,n\rangle}\!a^\dagger_ka^\dagger_la_ma_n+\rm{h. c.}\right),\nonumber\\
	\label{eq_HLHZ}
\end{eqnarray}
where ${a_k}$ and ${a^\dagger_k}$ are, respectively, the annihilation and creation operators for the $k$th KPO, and ${\hbar, K, p}$, and ${\Delta_k}$ are the reduced Planck constant, the Kerr coefficient, the pump amplitude, and the detuning frequency for the $k$th KPO, respectively.
This expression is obtained in a frame rotating at half the pump frequency, ${\omega_p/2}$, of the parametric drive and in the rotating-wave approximation~\cite{Dykman2012, Goto2016, Puri2017}.
In this work, we assume that $\Delta_k$ can be controlled individually.

$H_{\rm LHZ}$ corresponds to the LHZ energy.
In Eq.~(\ref{eq_HLHZ}), we introduce parameters, $\xi$ and $A$, to scale $H_{\rm LHZ}$ and the term including ${\left\{J_k\right\}}$, respectively, where $\xi$ has the dimension of frequency and ${A}$ is dimensionless.
${H_{\rm LHZ}}$ is small compared with ${H_{\rm KPO}}$, ${\xi\ll K}$, for accurate embedding~\cite{Goto2016, Puri2017}.
$A$ is chosen to reproduce ${E_{\rm LHZ}}$ (see Methods).
The first and second terms in ${H_{\rm LHZ}}$ physically mean local external drives at ${\omega_p/2}$~\cite{Goto2016a, Puri2017, Goto2018a} and four-body interactions, respectively.
Since the ancillary LHZ spins on the lower edge of the lattice are fixed to 1, as mentioned above, the corresponding KPOs can be replaced by classical driving~\cite{Puri2017}.
Thus their annihilation and creation operators in ${H_{\rm LHZ}}$ are replaced by classical driving amplitudes denoted by $\beta$ (see Appendix~\ref{sec_AppB}).

A solution of  the Ising problem is found from the ground state of the LHZ-QbM after adiabatic time evolution via bifurcations~\cite{Goto2016, Puri2017, Goto2019a}.
The sign of a quadrature amplitude ${x_k=\!\left(a_k+a^\dagger_k\right)/2}$ provides the LHZ spin ${\tilde{s}_k}$~\cite{Goto2016, Puri2017}, and the LHZ spins are transformed to Ising spins, giving the solution to the Ising problem.
Since the LHZ spins represent the relative signs of pairs of the Ising spins, a configuration ${\left\{\tilde{s}_k\right\}}$ corresponds to both ${\left\{s_i\right\}}$ and ${\left\{-s_i\right\}}$ simultaneously, which give an identical Ising energy.
Thus in mapping the LHZ spins to the Ising spins, we can fix one of the Ising spins, e.g., ${s_1=1}$.
The probabilities for the LHZ spins, hence for the Ising spins, can be evaluated from the wave function of the ground state.

In simulating a larger number of KPOs than before, we found that computational costs can be larger for the computation of the probabilities for the spins than that of time evolution.
These large costs come from the continuous degrees of freedom of the KPOs, which are finally projected to the binary variables.
We succeeded in calculating these probabilities within a reasonable computational time by noting that the Ising spins can be determined from only ${(N-1)}$ KPOs~\cite{Lechner2015} and thus tracing out the others (see Methods).
For instance, only the KPOs ${1, 2,}$ and $3$ are readout in the case where ${N=4}$, as shown in Fig.~\ref{fig_KPONet}.

Before showing simulation results, we evaluate the ground state by an analytical method, in which we find inhomogeneity in the KPOs.
Since this inhomogeneity will degrade the accuracy of the solutions, we propose a method to correct the inhomogeneity.
\\

\noindent{\bf Inhomogeneous photon numbers in the LHZ-QbM}\\
We examine the mean photon numbers in the ground state of $H$ by using a variational method.
Here, we assume a small $\xi$ satisfying ${\xi/K\ll1}$.
Although the terms yielded by the four-body interactions in Eq.~(\ref{eq_HLHZ}) seem intractable, these terms are simplified when the four-body constraints hold (see Methods).
As a result, the mean photon number in the $k$th KPO can be approximated by
\begin{eqnarray}
	\left\langle a^\dagger_ka_k\right\rangle\simeq\frac{p-\Delta_k}{K}+\frac{\xi}{K}\!\left(\frac{A}{\alpha_0}J_k\tilde{s}_k+\alpha^2_0Cz_k\right),\label{eq_PhotNum}
\end{eqnarray}
where ${\alpha_0=\!\left[\left(p-\bar{\Delta}\right)/K\right]^{1/2}}$, and ${\bar{\Delta}}$ is the average of ${\Delta_k}$ over the KPOs.
${z_k}$ is the number of the four-body interactions connected to the $k$th KPO.
The first term ${\left(p-\Delta_k\right)/K}$ is the mean photon number without ${H_{\rm LHZ}}$, and the second term indicates inhomogeneity induced by ${H_{\rm LHZ}}$.
[See Methods for the detailed derivation of Eq.~(\ref{eq_PhotNum}).]

In Eq.~(\ref{eq_PhotNum}), the terms proportional to ${J_k\tilde{s}_k}$ and ${Cz_k}$ are caused by the local external drives and the four-body interactions, respectively.
Because the term proportional to ${Cz_k}$ typically becomes dominant under the condition for satisfying the four-body constraints, we focus on this term in the following.
In general, ${z_k}$ takes the values of ${1, 2, 3}$, and $4$ as can be seen from Fig.~\ref{fig_KPONet}.
Hence the variation in the mean photon number due to the four-body interactions can become smaller for KPOs near to the edge of the network while larger around its center.

This inhomogeneity due to the four-body interactions effectively changes the local drive in each KPO and consequently prevents the local external drives in Eq.~(\ref{eq_HLHZ}) from accurately implementing ${\left\{J_k\right\}}$, which results in low solution accuracy.
The solution accuracy is thus expected to be improved by reducing the inhomogeneity in the photon numbers.
Here we propose a method to reduce the inhomogeneity without measurement.
\\

\noindent{\bf Correcting the inhomogeneous photon numbers}\\
We reduce the inhomogeneity in the mean photon numbers by setting $\Delta_k$ as
\begin{eqnarray}
	\Delta_k=\Delta+\frac{p-\Delta}{K}\xi Cz_k,\label{eq_Delta}
\end{eqnarray}
for ${p-\Delta>0}$, where $\Delta$ is a common detuning frequency.
$\Delta_k$ in Eq.~(\ref{eq_Delta}) determines ${\bar{\Delta}}$, hence ${\alpha_0}$ as $\alpha^2_0=\!\left[(p-\Delta)/K\right]\!\left(1-\xi C\bar{z}/K\right)$, where $\bar{z}$ is the average of $z_k$.
Thus, within the first order approximation in ${\xi/K}$, $\alpha^2_0$ in Eq.~(\ref{eq_PhotNum}) can be replaced by $(p-\Delta)/K$.
Substituting Eq.~(\ref{eq_Delta}) into the first term in Eq.~(\ref{eq_PhotNum}), we obtain homogeneous mean photon numbers:
\begin{eqnarray}
	\left\langle a^\dagger_ka_k\right\rangle\simeq\frac{p-\Delta}{K},\label{eq_PhotNumCorr}
\end{eqnarray}
where the second term in Eq.~(\ref{eq_Delta}) has  canceled the term proportional to $C$ in Eq.~(\ref{eq_PhotNum}), and  the term proportional to ${J_k\tilde{s}_k}$ has been dropped.

The correction by Eq.~(\ref{eq_Delta}) is possible without knowing the solutions to the Ising problems, because the term proportional to $C$ in Eq.~(\ref{eq_PhotNum}) is independent of ${\left\{J_k\right\}}$ or ${\left\{\tilde{s}_k\right\}}$.
The parameters are therefore scheduled in advance, and no measurement of the states is necessary during the time evolution.

In the following, we simulate the time evolution of the LHZ-QbM shown in Fig.~\ref{fig_KPONet} by numerically solving the Schr\"{o}dinger equation with $H$.
(See Methods for details.)
${N=4}$, hence ${L=6}$, is chosen because this network is the smallest among those where the KPOs are unequal.
We first apply this LHZ-QbM to the Ising problem with uniform interactions to check the inhomogeneity, and then with random interactions in order to evaluate the performance.
\\

\begin{figure}[b]
	\includegraphics[width=7cm]{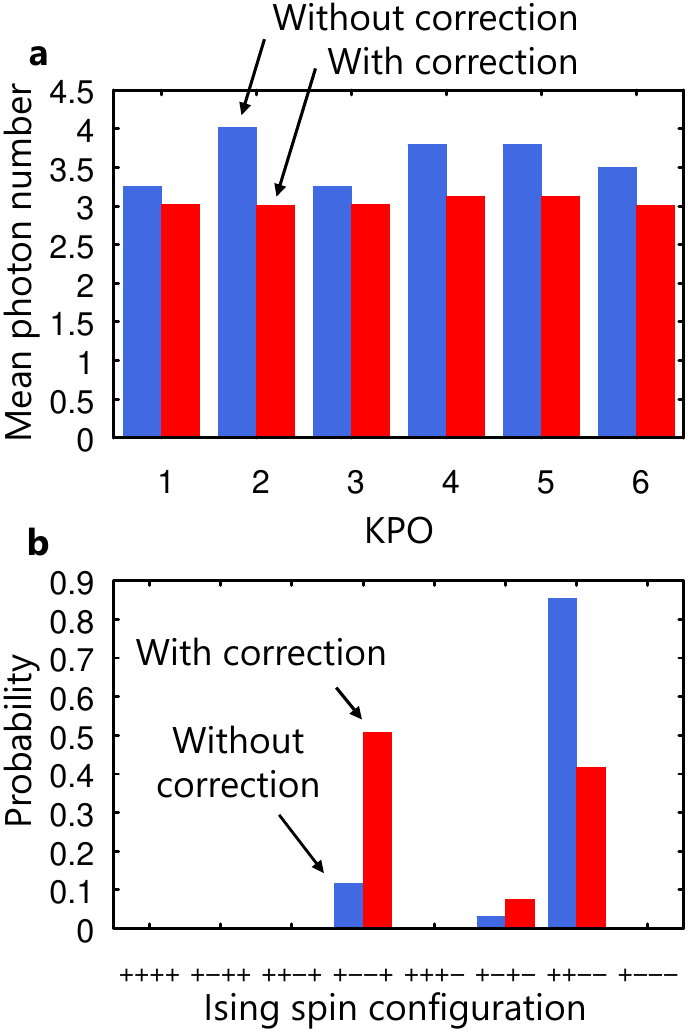}%
	\caption{{\bf Results for the four-spin Ising problem with uniform antiferromagnetic interactions.}
		{\bf a} Mean photon number in each KPO for ${C=0.3}$ and ${\xi/K=0.3}$.
		Here, ${C=0.3}$ is chosen because this satisfies the condition, ${C>1/6}$, for satisfying the four-body constraints (see Appendix~\ref{sec_AppA}).
		{\bf b} Probabilities for the configurations of the Ising spins, ${\left\{s_i\right\}}$.
		Only the configurations with ${s_1=1}$ are shown for the reason mentioned in the main text.\label{fig_Uni}}
\end{figure}

\begin{figure*}
	\begin{center}
	\includegraphics[width=17cm]{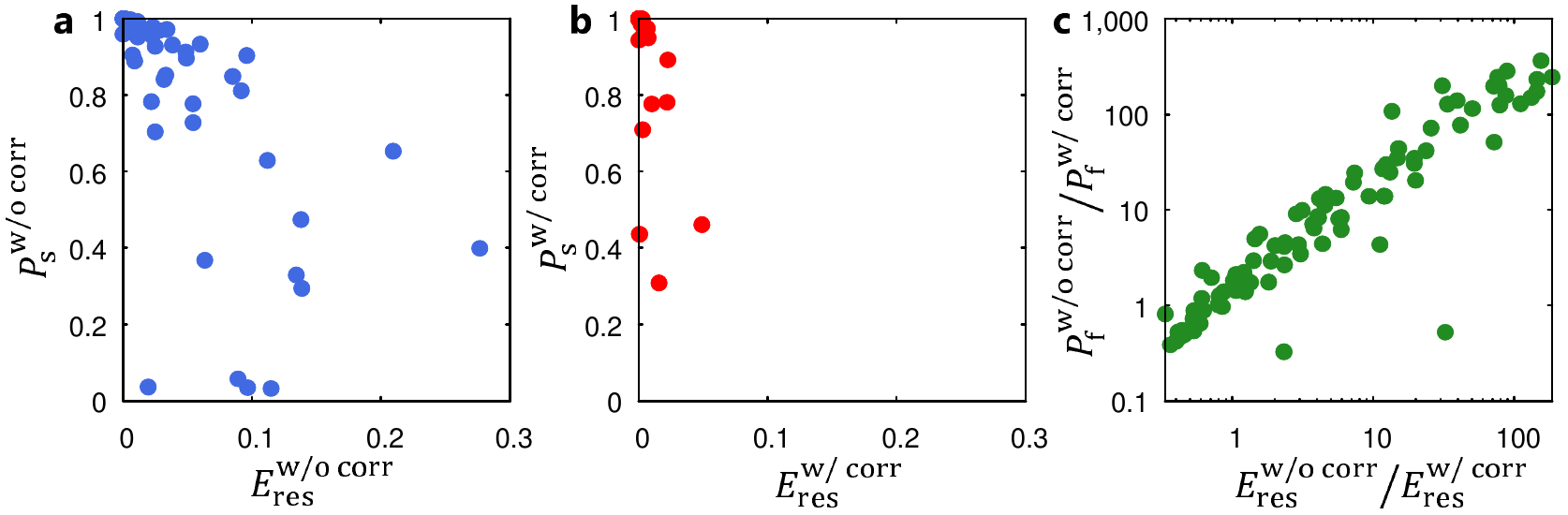}%
	\caption{{\bf Results for 100 random instances of the four-spin Ising problem.}
		{\bf a} Distributions of success probabilities ${P^{\rm w/o\hspace{0.2em}corr}_{\rm s}}$ and residual energies ${E^{\rm w/o\hspace{0.2em}corr}_{\rm res}}$ without the correction.
		{\bf b} Corresponding results with the correction, ${P^{\rm w/ corr}_{\rm s}}$ and ${E^{\rm w/ corr}_{\rm res}}$.
		{\bf c} Distributions of improvement rates defined as ${P^{\rm w/o\hspace{0.2em}corr}_{\rm f}/P^{\rm w/ corr}_{\rm f}}$ and ${E^{\rm w/o\hspace{0.2em}corr}_{\rm res}/E^{\rm w/ corr}_{\rm res}}$.\label{fig_SProbOpt}}
	\end{center}
\end{figure*}

\noindent{\bf Simulation results for uniform antiferromagnetic interactions}\\
Here we solve the Ising problem with uniform all-to-all connected antiferromagnetic interactions~\cite{Marandi2014}, which is expressed by the same negative ${J}$ for all ${\left\{J_k\right\}}$.
In this work, we normalize the ${\{J_k\}}$ such that ${\sum^L_{k=1}\left|J_k\right|=1}$.
 (This normalization does not change the Ising problem.)
Thus in the present case, ${J_k=-1/L}$.

Figure~\ref{fig_Uni}a shows the mean photon number in each KPO after time evolution, where we compare the results for uniform ${\Delta}$ (without the correction) and modified ${\Delta_k}$ in Eq.~(\ref{eq_Delta}) (with the correction).
For uniform $\Delta$, the mean photon numbers increase depending on ${z_k}$ (${z_1=z_3=z_6=1}$, ${z_4=z_5=2}$, and ${z_2=3}$ as can be seen from Fig.~\ref{fig_KPONet}), which are consistent with Eq.~(\ref{eq_PhotNum}).
Since ${\left\{J_k\right\}}$ is uniform, this inhomogeneity originates from the four-body interactions.

For modified ${\Delta_k}$, on the other hand, all the mean photon numbers are nearly equal to 3, which coincides with the value predicted by Eq.~(\ref{eq_PhotNumCorr}).
This result shows that the photon numbers can be made almost homogeneous by using ${\Delta_k}$ in Eq.~(\ref{eq_Delta}) as expected.

Figure~\ref{fig_Uni}b shows the probabilities for the Ising spin configurations.
Although the probabilities are finite only for the three degenerate ground states, namely two spins in $1$ and the others ${-1}$, the probabilities for these three configurations are not equal.
In particular, uniform $\Delta$ results in the much higher probability for ${\{++--\}}$ than the other configurations.
The reason for this can be explained as follows.
The configuration ${\{++--\}}$ is mapped to the LHZ spins of ${\tilde{s}_1=\tilde{s}_3=1}$ and the others in ${-1}$.
While the negative ${J_k}$ favors ${\tilde{s}_k= -1}$, the four-body constraints lead to ${\tilde{s}_1=\tilde{s}_3=1}$.
Here ${J_1}$ and ${J_3}$ become effectively small because of the relatively small mean photon numbers in KPOs 1 and 3, as shown in Fig.~\ref{fig_Uni}a, which is caused by the four-body interactions.
Thus the probability for this configuration becomes particularly high compared to the others.

When ${\Delta_k}$ in Eq.~(\ref{eq_Delta}) are used, the probabilities are similar for two of the ground states, ${\{+--+\}}$ and ${\{++--\}}$, as shown in Fig.~\ref{fig_Uni}b.
Thus, the bias to one ground state ${\{++--\}}$ observed for uniform $\Delta$ are suppressed by reducing the inhomogeneity in the photon numbers.

\noindent{\bf Simulation results for random interactions}\\
Finally we evaluate the performance of the LHZ-QbM for the Ising problem with all-to-all connected random interactions, generating ${\left\{J_{ij}\right\}}$ uniformly from ${\{-1, -0.99, -0.98, \cdots, 1\}}$ for 100 instances~\cite{Goto2016, Goto2018a, Goto2019a}, and normalizing them such that ${\sum^L_{k=1}\left|J_k\right|=1}$, as mentioned above.
The performance is measured by a success probability ${P_{\rm s}}$ and a residual energy ${E_{\rm res}}$.
The success probability is defined as a probability for obtaining ${\left\{s_i\right\}}$ minimizing ${E_{\rm Ising}}$.
The residual energy is given by the expectation value of ${E_{\rm Ising}}$ subtracted by the minimum ${E_{\rm Ising}}$~\cite{Santoro2002, Goto2016, Goto2019a}.
A lower residual energy indicates higher accuracy, and takes its minimum of 0 when the success probability is 1.

Figures~\ref{fig_SProbOpt}a and~\ref{fig_SProbOpt}b show the distributions of ${P_{\rm s}}$ and ${E_{\rm res}}$ without and with the correction, respectively.
In each case, the values of ${(C, \xi/K)}$ are set to ${(0.3, 0.3)}$ and ${(0.4, 0.6)}$, in order to maximize the success probabilities averaged over the instances (see Appendix~\ref{sec_AppD}).
Without the correction, for several instances the success probabilities are nearly 0.
With the correction, in contrast, the success probabilities are at least higher than $0.3$ and close to 1 for most instances.
With the nonzero success probability, repeated use of this LHZ-QbM gives a correct solution with probability rapidly approaching 1.
The residual energies are also substantially lowered by the correction, which means that obtained solutions become more accurate.

In Fig.~\ref{fig_SProbOpt}c we show improvement rates of a failure probability ${P_{\rm f}=1-P_{\rm s}}$ and ${E_{\rm res}}$ in each instance, where an improvement rate is defined as the ratio of the value without the correction to that with it.
Figure~\ref{fig_SProbOpt}c shows that both ${P_{\rm f}}$ and ${E_{\rm res}}$ are improved by one or two orders of magnitude in many instances, with the factors of up to ${P^{\rm w/o\hspace{0.2em}corr}_{\rm f}/P^{\rm w/ corr}_{\rm f}=364}$ and ${E^{\rm w/o\hspace{0.2em}corr}_{\rm res}/E^{\rm w/ corr}_{\rm res}=185}$, indicating dramatic improvements by the proposed correction.

\section*{DISCUSSION}
In the LHZ-QbM, i.e.~a KPO network for adiabatic quantum computation in the LHZ scheme, we have shown that inhomogeneous photon numbers due to four-body interactions degrade its performance, and furthermore that the inhomogeneity can be suppressed by controlling detunings, which can dramatically improve the performance.
This method does not need to refer to the states of the KPOs during adiabatic time evolution, offering a simple operation.
This method can be used regardless of the number of KPOs in an LHZ-QbM, thus allowing its large-scale implementation.

While we have modified only the detunings in the present work, similar corrections are possible by setting Kerr coefficients or pump amplitudes.
In implementations with superconducting circuits, these three parameters can be controlled experimentally through parameters characterizing these circuits~\cite{Puri2017, Goto2019a}.

In the present study, we have assumed sufficiently small ${H_{\rm LHZ}}$ compared to ${H_{\rm KPO}}$, and based on this assumption, we have determined the detunings to correct the inhomogeneity.
From the simulation results, however, we have found that the optimal value, $\xi/K=0.6$, is rather large (see Appendix~\ref{sec_AppD}).
This large optimal $\xi/K$ may be because the time evolution becomes more adiabatic, that is, the larger ${H_{\rm LHZ}}$ widens gaps between the energy levels of ground states and excited states, and prevents transitions between these states during the time evolution~\cite{Albash2018}.
These results imply further possible improvement by increasing ${\xi/K}$ together with more elaborate control of detunings than the present analytic ones.

\section*{METHODS}
\begin{figure}
	\includegraphics[width=7cm]{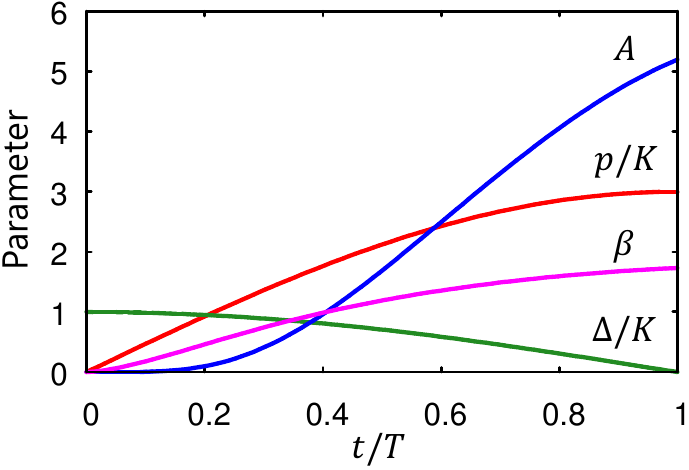}%
	\caption{{\bf Time dependence of parameters.}
		The others, ${K, \xi}$, and $C$, are set to be constant~\cite{Puri2017, Goto2019a}, and positive $K$ is assumed~\cite{Goto2016, Goto2019a}.
		(For negative $K$, the same results can be obtained by changing the signs of ${p, \Delta}$, and $\xi$~\cite{Goto2016, Goto2019a}.
 		Negative $K$ has been used in Ref.~\cite{Puri2017}.)\label{fig_Param}}
\end{figure}

\noindent{\bf Simulation of the LHZ-QbM}\\
Each KPO is initialized to a vacuum, and the parameters in $H$ are set such that the vacuum is its ground state.
For the initial parameters ${p/K=A=\beta=0}$ and ${\Delta/K>0}$, the vacuum is the ground state if ${\xi C/K\leq 1}$, which is valid within an estimation by a variational method~\cite{Goto2016, Goto2018a} (see Appendix~\ref{sec_AppC}).
Time evolution is then started, and the parameters are slowly varied to induce bifurcations.
During the evolution, the state is expected to remain in the ground state.
Figure~\ref{fig_Param} shows the parameters as a function of time~\cite{Goto2016, Goto2018a}, where the time interval for the evolution is ${T=500/K}$.
To reproduce ${E_{\rm LHZ}}$, ${A}$ and ${\beta}$ for ${t\simeq T}$ are chosen such that ${A\simeq\alpha^3}$ and ${\beta\simeq\alpha}$, where ${\alpha=[(p-\Delta)/K]^{1/2}}$.

After the bifurcations, the ground state approximately becomes an $L$-mode coherent state ${\left|\alpha_1\right\rangle\cdots\!\left|\alpha_L\right\rangle}$~\cite{Goto2016, Puri2017, Goto2019a}, where the coherent state ${\left|\alpha_k\right\rangle}$ satisfies $a_k\!\left|\alpha_k\right\rangle=\alpha_k\!\left|\alpha_k\right\rangle$~\cite{Leonhardt1997}.
The signs of ${\left\{\alpha_k\right\}}$ are expected to give the LHZ spins minimizing $E_{\rm LHZ}$~\cite{Puri2017}.

The time evolution is simulated by numerically solving the Schr\"{o}dinger equation, where $H$ and the state ${\left|\psi\right\rangle}$ are represented in the photon number basis with the largest photon number 12 truncating the Hilbert space~\cite{Goto2016, Goto2018a}.

After the time evolution, the probabilities for the LHZ spins can be formulated using eigenstates of ${x_k}$ as~\cite{Goto2016, Goto2018a, Leonhardt1997}
\begin{eqnarray}
	P\!\left(\tilde{s}_1,\cdots,\tilde{s}_L\right)&=&\mathrm{Tr}\!\left[\!\left|\psi\right\rangle\!\left\langle\psi\right|\!\prod^L_{k=1}M_k\!\left(\tilde{s}_k\right)\right],\label{eq_ProbLHZ}\\
	M_k\!\left(\pm1\right)&=&\!\int_{x_k\gtrless0}\!\mathrm{d}x_k\!\left|x_k\right\rangle\!\left\langle x_k\right|.
\end{eqnarray}
However we found that their calculation becomes the largest in the simulation for large $L$.
Then, noting that $(N-1)$ LHZ spins can determine the Ising spins~\cite{Lechner2015}, we trace out the other KPOs in Eq.~(\ref{eq_ProbLHZ}) without corresponding ${M_k\!\left(\tilde{s}_k\right)}$ and obtain the probabilities within reasonable computational costs.
\\

\noindent{\bf Inhomogeneity in the KPOs in the LHZ-QbM}\\
We evaluate the ground state using a variational method and assuming the bifurcations, small ${H_{\rm LHZ}}$, and the four-body constraints.
We employ ${\left|\alpha_1\right\rangle\cdots\!\left|\alpha_L\right\rangle}$ as a trial wave function~\cite{Goto2016, Goto2018a}, and minimize the expectation value ${\langle H\rangle}$ with respect to ${\left\{\alpha_k, \alpha^*_k\right\}}$, where ${\alpha^*_k}$ is the complex conjugate of  ${\alpha_k}$.
The following nonlinear equations are then yielded
\begin{eqnarray}
	&&K\alpha^*_k\alpha^2_k-p\alpha^*_k+\Delta_k\alpha_k-\xi\!\left(AJ_k+C\!\sum_{\langle l,m,n\rangle}\!\alpha^*_l\alpha_m\alpha_n\right)=0.\nonumber\\
	\label{eq_Vari}
\end{eqnarray}
By assuming sufficiently small $\xi$, Eq.~(\ref{eq_Vari}) is approximately solved within the first order in $\xi$ as
\begin{eqnarray}
	\alpha_k&\simeq&\alpha_{0k}\tilde{s}_k+\!\frac{\xi}{2\alpha^2_0K}\!\left(AJ_k+\alpha^3_0C\!\sum_{\langle l,m,n\rangle}\!\tilde{s}_l\tilde{s}_m\tilde{s}_n\right),\label{eq_Alpha1}
\end{eqnarray}
where ${\alpha_{0k}=\!\left[\left(p-\Delta_k\right)/K\right]^{1/2}}$, and ${\alpha_{0k}\tilde{s}_k}$ is the solution of Eq.~(\ref{eq_Vari}) for ${\xi=0}$ after the bifurcation ${p-\Delta_k>0}$.
Here ${\Delta_k-\bar{\Delta}}$ is assumed to be the first order in $\xi$ [which is valid in Eq.~(\ref{eq_Delta})].

The term proportional to $C$ in Eq.~(\ref{eq_Alpha1}) can be simplified under the four-body constraints ${\tilde{s}_k\tilde{s}_l\tilde{s}_m\tilde{s}_n=1}$ as follows.
Since ${\tilde{s}_l\tilde{s}_m\tilde{s}_n=\tilde{s}_k}$ holds, the summation in Eq.~(\ref{eq_Alpha1}) reduces to the number of connected four-body interactions ${z_k}$, resulting in
\begin{eqnarray}
	\alpha_k\simeq\alpha_{0k}\tilde{s}_k+\!\frac{\xi}{2\alpha^2_0K}\!\left(AJ_k+\alpha^3_0C\tilde{s}_kz_k\right).\label{eq_Alpha2}
\end{eqnarray}
This expression becomes accurate for $\xi/K\ll2/\!\left(\!\left|J_k\right|+Cz_k\right)\sim1$.
Equation~(\ref{eq_Alpha2}) shows that ${H_{\rm LHZ}}$ affects the amplitude ${\left|\alpha_k\right|}$.
Its square ${\left|\alpha_k\right|^2}$ gives the mean photon number in Eq.~(\ref{eq_PhotNum}) in this approximation.

\begin{acknowledgements}
This work was supported by JST ERATO (Grant No. JPMJER1601).
\end{acknowledgements}

\appendix

\section{Condition on $C$ for satisfying the four-body constraints}\label{sec_AppA}
We derive a formula that expresses the condition on $C$ for satisfying the four-body constraints for ${\left\{\tilde{s}_k\right\}}$ minimizing ${E_{\rm LHZ}}$.
We then show that ${C>1}$ is a sufficient condition under the normalization condition ${\sum^L_{k=1}\left|J_k\right|=1}$.

First of all, it is notable that the second term in ${E_{\rm LHZ}}$, ${-C\!\sum_{\langle k, l, m, n\rangle}\!\tilde{s}_k\tilde{s}_l\tilde{s}_m\tilde{s}_n}$, is simply expressed by $-C(L-N+1-2b)$, where ${(L-N+1)}$ is the number of all the constraints and $b$ denotes the number of broken constraints.
Thus in general, minimum ${E_{\rm LHZ}}$ with $b$ broken constraints, which is denoted by ${E^{(b)}_{\rm LHZ}}$, is expressed as
\begin{eqnarray}
	E^{(b)}_{\rm LHZ}=-\sum^L_{k=1}J_k\tilde{s}^{(b)}_k-C(L-N+1-2b),\label{eq_EB}
\end{eqnarray}
where ${\left\{\tilde{s}^{(b)}_k\right\}}$ is the spin configuration minimizing ${E_{\rm LHZ}}$ with $b$ broken constraints.
By using these notations, the condition for satisfying the constraints is written as ${E^{(0)}_{\rm LHZ}<E^{(b)}_{\rm LHZ}}$ for all ${b\geq1}$.
Thus, using Eq.~(\ref{eq_EB}), we have the following inequality:
\begin{eqnarray}
	C>\frac{1}{2b}\left[-\sum^L_{k=1}J_k\tilde{s}^{(0)}_k+\sum^L_{k=1}J_k\tilde{s}^{(b)}_k\right].\label{eq_CEno}
\end{eqnarray}
Note that the right-hand side in Eq.~(\ref{eq_CEno}) offers a tight lower bound for $C$ to satisfy the constraints for an arbitrary instance ${\left\{J_k\right\}}$.
In the case of the uniform antiferromagnetic interaction in the main text (${L=6}$ and ${J_k=-1/6}$), the values of ${-\sum^6_{k=1}J_k\tilde{s}^{(b)}_k}$ for ${b=0,1,2,3}$ are ${-1/3, -2/3, -1}$, and ${-2/3}$, respectively, and thus Eq.~(\ref{eq_CEno}) gives ${C>1/6}$.

Next, we roughly evaluate a condition on $C$ valid for any instance ${\left\{J_k\right\}}$.
The right-hand side of Eq.~(\ref{eq_CEno}) can be estimated as
\begin{eqnarray}
	-\frac{1}{2b}\sum^L_{k=1}J_k\left[\tilde{s}^{(0)}_k-\tilde{s}^{(b)}_k\right]\leq\sum^L_{k=1}\left|J_k\right|=1,\label{eq_Upper}
\end{eqnarray}
where we have used ${b\geq1, \left|\tilde{s}^{(0)}_k-\tilde{s}^{(b)}_k\right|\leq2}$, and the normalization condition ${\sum^L_{k=1}\left|J_k\right|=1}$.
Equation~(\ref{eq_Upper}) shows that the right-hand side of Eq.~(\ref{eq_CEno}) is at most 1, indicating that ${C>1}$ is a sufficient condition for satisfying the four-body constraints.

\section{${H_{\rm LHZ}}$ for 6 KPOs}\label{sec_AppB}
Here we present ${H_{\rm LHZ}}$ more explicitly.
The oscillators denoted by ``Fixed'' in the lowest row in Fig.~\ref{fig_KPONet} are assumed to be in a coherent state ${\left|\beta\right\rangle}$ with ${\beta>0}$~\cite{Puri2017}.
As these oscillators do not bifurcate and $\beta$ can be controlled by external fields, we replace corresponding annihilation and creation operators by $\beta$ and regard $\beta$ as a parameter.
As a result, the LHZ part of the Hamiltonian is written as
\begin{eqnarray}
	H_{\rm LHZ}&=&-\hbar\xi\Bigg[A\!\sum^6_{k=1}\!J_ka^\dagger_k\\
	&&+C\!\left(\beta a^\dagger_1a^\dagger_2a_4+\beta a^\dagger_2a^\dagger_3a_5+a^\dagger_2a^\dagger_4a_5a_6\right)+\rm{h. c.}\Bigg].\nonumber
\end{eqnarray}

\section{The ground state of the initial Hamiltonian}\label{sec_AppC}
We consider a condition for the vacuum to be the ground state of $H$ at initial time for ${L=6}$.
As we set ${p/K=A=\beta=0}$ and ${\Delta_k=\Delta}$ at ${t=0}$, $H$ is written as
\begin{eqnarray}
	H=\hbar\!\sum^6_{k=1}\!\left(\frac{K}{2}a^{\dagger2}_ka^2_k+\Delta a^{\dagger}_ka_k\right)-\hbar\xi C\!\left(a^\dagger_2a^\dagger_4a_5a_6+{\rm h.c.}\right).\nonumber\\
\end{eqnarray}

\begin{figure*}
	\begin{center}
		\includegraphics[width=13.3cm]{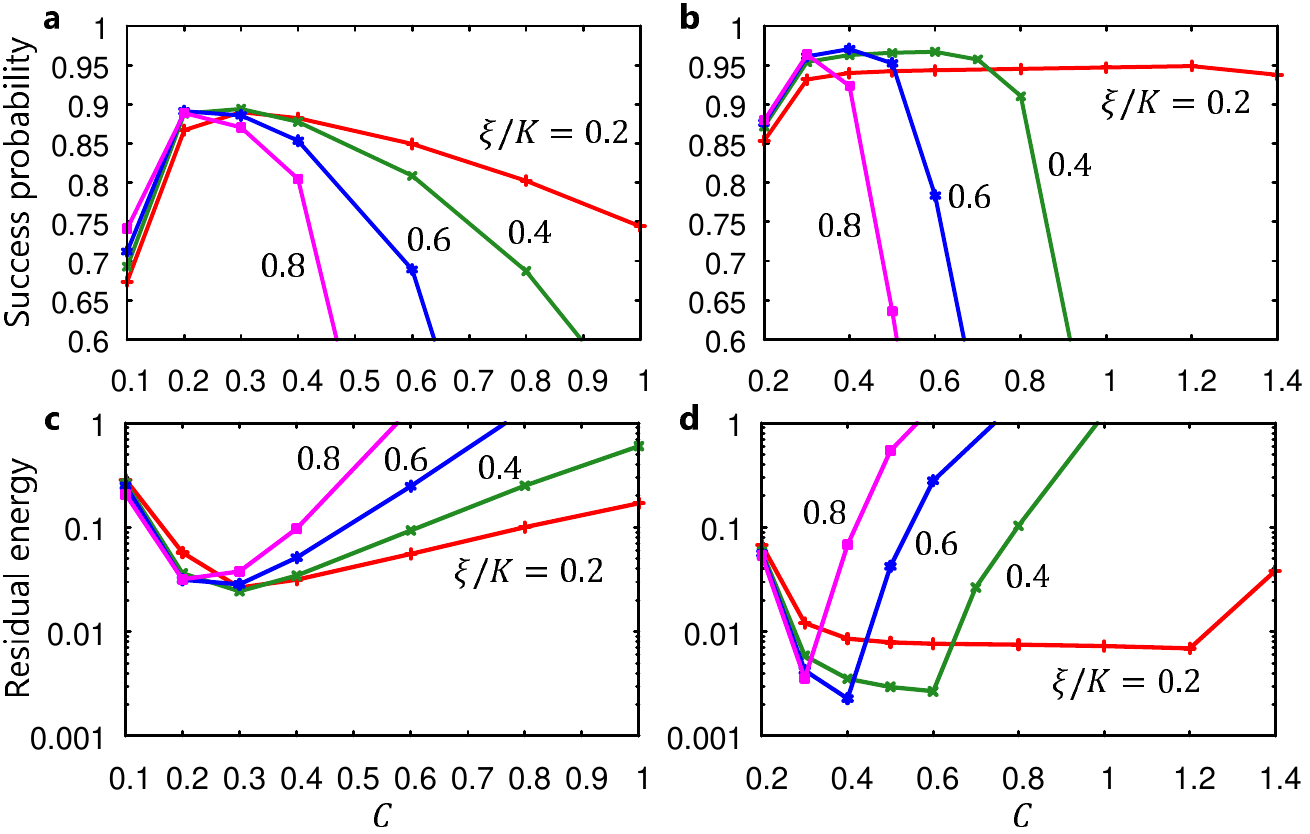}%
		\caption{{\bf Average performance for the 100 random instances.}
			{\bf a} Success probabilities averaged over the random instances discussed in the main text without the correction.
			{\bf b} Corresponding results with the correction.
			{\bf c} Average residual energies without the correction.
			{\bf d} Corresponding results with the correction.\label{fig_FProb}}
	\end{center}
\end{figure*}

We evaluate the ground state by the variational method, using the coherent state ${\left|\alpha_1\right\rangle\cdots\left|\alpha_6\right\rangle}$ as a trial wave function~\cite{Goto2016, Goto2018a}.
${\partial\langle H\rangle/\partial\alpha^*_k=0}$ gives, for ${k=1, 3}$, \begin{eqnarray}
	K\alpha^*_k\alpha^2_k+\Delta \alpha_k=0, \end{eqnarray} and for the others, \begin{eqnarray}
	K\alpha^*_2\alpha^2_2+\Delta\alpha_2-\xi C\alpha^*_4\alpha_5\alpha_6&=&0,\label{eq_Alp2}\\
	K\alpha^*_4\alpha^2_4+\Delta\alpha_4-\xi C\alpha^*_2\alpha_5\alpha_6&=&0,\\
	K\alpha^*_5\alpha^2_5+\Delta\alpha_5-\xi C\alpha_2\alpha_4\alpha^*_6&=&0,\\
	K\alpha^*_6\alpha^2_6+\Delta\alpha_6-\xi C\alpha_2\alpha_4\alpha^*_5&=&0.\label{eq_Alp6}
\end{eqnarray}
KPOs 1 and 3 are decoupled and their ground states are the vacuums since ${K>0}$ and ${\Delta>0}$.
Equations~(\ref{eq_Alp2})--(\ref{eq_Alp6}) can be transformed to \begin{eqnarray}
	\left[\!\prod_{k}\!\left(K\!\left|\alpha_k\right|^2+\Delta\right)-(\xi C)^4\!\prod_{k}\left|\alpha_k\right|^2\right]\!\prod_{k}\alpha_k=0,
\end{eqnarray}
where the products are over ${k=2,4,5,6}$.
If ${K\geq\xi C}$, the quantity in the square bracket is always positive and thus ${\alpha_2\alpha_4\alpha_5\alpha_6}$ must be zero.
${\alpha_2\alpha_4\alpha_5\alpha_6=0}$ and Eqs.~(\ref{eq_Alp2})--(\ref{eq_Alp6}) result in ${\alpha_2=\alpha_4=\alpha_5=\alpha_6=0}$.
Around ${\alpha_k=0}$, ${\langle H\rangle}$ is approximated by ${\langle H\rangle\simeq\hbar\sum^6_{k=1}\Delta\left|\alpha_k\right|^2}$, which shows that ${\alpha_k=0}$ corresponds to the minimum.
This result means that if ${\xi C/K\leq1}$ then the initial ground state is the vacuum within the estimation by the variational method.

\section{Average success probabilities and residual energies}\label{sec_AppD}
Figures~\ref{fig_FProb}a (without the correction) and \ref{fig_FProb}b (with the correction) show the success probabilities averaged over the 100 random instances in the main text as functions of $C$ and $\xi$.
By the proposed correction, the highest average success probability is increased from $89.5$\% at ${(C, \xi/K)=(0.3, 0.3)}$ to $97.1$\% at ${(0.4, 0.6)}$, which are used for Fig.~\ref{fig_SProbOpt}.
The average residual energies are shown in Figs.~\ref{fig_FProb}c (without the correction) and \ref{fig_FProb}d (with the correction), which are decreased by one order of magnitude by the correction.

The dependences of the average success probabilities on ${C}$ are understood as follows.
(Those of the residual energies can be understood in the same way.)
We first explain the case without the correction shown in Fig.~\ref{fig_FProb}a.
As a function of $C$, the average success probability shows a maximum.
The low success probabilities for ${C\ll1}$ are because the four-body constraints are frequently broken.
With increasing $C$, the inhomogeneity in the photon numbers due to the four-body interactions is increased and degrades the success probabilities.
Thus the average success probabilities are the highest at an intermediate $C$.

For the case with the correction shown in Fig.~\ref{fig_FProb}b, the average success probability also shows a maximum as a function of $C$.
For small $C$, the success probabilities are low, again, owing to violated four-body constraints as mentioned above.
For intermediate $C$, the correction can suppress the effects of the inhomogeneity in the photon numbers, unlike the above case, leading to higher performance.
For larger $C$, however, the success probabilities suddenly decrease, because the estimation given by Eq.~(\ref{eq_PhotNum}) is no longer valid for these ${(C, \xi/K)}$, and the resultant large modulation of detunings according to Eq.~(\ref{eq_Delta}) rather degrades the performance.


\begin{thebibliography}{99}
\bibitem{Barahona1982}Barahona, F. On the computational complexity of Ising spin glass models. {\it J. Phys. A: Math. Gen.} {\bf 15,} 3241 (1982).
\bibitem{Lucas2014}Lucas, A. Ising formulations of many NP problems. {\it Front. Phys.} {\bf 2,} 5 (2014).
\bibitem{Barahona1988}Barahona, F., Gr\"{o}tschel, M., J\"{u}nger, M. \& Reinelt, G. An application of combinatorial optimization to statistical physics and circuit layout design. {\it Oper. Res.} {\bf 36,} 493 (1988).
\bibitem{Perdomo-Ortiz2012}Perdomo-Ortiz, A., Dickson, N., Drew-Brook, M., Rose, G. \& Aspuru-Guzik, A. Finding low-energy conformations of lattice protein models by quantum annealing. {\it Sci. Rep.} {\bf 2,} 571 (2012).
\bibitem{Li2018}Li, R. Y., Di Felice, R., Rohs, R. \& Lidar, D. A. Quantum annealing versus classical machine learning applied to a simplified computational biology problem. {\it npj Quant. Inform.} {\bf 4,} 14 (2018).
\bibitem{Rosenberg2016}Rosenberg, G., Haghnegahdar, P., Goddard, P., Carr, P., Wu, K. \& L\'{o}pez de Prado, M. Solving the optimal trading trajectory problem using a quantum annealer. {\it IEEE J. Sel. Top. Sig. Process.} {\bf 10,} 1053 (2016).
\bibitem{Kadowaki1998}Kadowaki, T. \& Nishimori, H. Quantum annealing in the transverse Ising model. {\it Phys. Rev. E} {\bf 58,} 5355 (1998).
\bibitem{Das2008}Das, A. \& Chakrabarti, B. K. Colloquium: Quantum annealing and analog quantum computation. {\it Rev. Mod. Phys.} {\bf 80,} 1061 (2008).
\bibitem{Farhi2000}Farhi, E., Goldstone, J. Gutmann, S. \& Sipser, M. Quantum computation by adiabatic evolution. Preprint at arXiv:quant-ph/0001106 (2000).
\bibitem{Farhi2001}Farhi, E., Goldstone, J., Gutmann, S., Lapan, J., Lundgren, A. \& Preda, D. A quantum adiabatic evolution algorithm applied to random instances of an NP-complete problem. {\it Science} {\bf 292,} 472 (2001).
\bibitem{Albash2018}Albash, T. \& Lidar, D. A. Adiabatic quantum computation. {\it Rev. Mod. Phys.} {\bf 90,} 015002 (2018).
\bibitem{Johnson2011}Johnson, M. W., Amin, M. H. S., Gildert, S., Lanting, T., Hamze, F., Dickson, N., Harris, R., Berkley, A. J., Johansson, J., Bunyk, P., Chapple, E. M., Enderud, C., Hilton, J. P., Karimi, K., Ladizinsky, E., Ladizinsky, N., Oh, T., Perminov, I., Rich, C., Thom, M. C., Tolkacheva, E., Truncik, C. J. S., Uchaikin, S., Wang, J., Wilson, B. \& Rose, G. Quantum annealing with manufactured spins. {\it Nature} {\bf 473,} 194 (2011).
\bibitem{Wang2013}Wang, Z., Marandi, A., Wen, K., Byer, R. L. \& Yamamoto, Y. Coherent Ising machine based on degenerate optical parametric oscillators. {\it Phys. Rev. A} {\bf 88,} 063853 (2013).
\bibitem{Marandi2014}Marandi, A., Wang, Z., Takata, K., Byer, R. L. \& Yamamoto, Y. Network of time-multiplexed optical parametric oscillators as a coherent Ising machine. {\it Nat. Photonics} {\bf 8,} 937 (2014).
\bibitem{Leleu2017}Leleu, T., Yamamoto, Y., Utsunomiya, S. \& Aihara, K. Combinatorial optimization using dynamical phase transitions in driven-dissipative systems. {\it Phys. Rev. E} {\bf 95,} 022118 (2017). 
\bibitem{Yamamoto2017}Yamamoto, Y., Aihara, K., Leleu, T., Kawarabayashi, K., Kako, S., Fejer, M., Inoue, K. \& Takesue, H. Coherent Ising machines--optical neural networks operating at the quantum limit. {\it npj Quant. Inform.} {\bf 3,} 49 (2017).
\bibitem{Kirkpatrick1983}Kirkpatrick, S., Gelatt, C. D. \& Vecchi, M. P. Optimization by simulated annealing. {\it Science} {\bf 220,} 671 (1983).
\bibitem{Yamaoka2016}Yamaoka, M., Yoshimura, C.,Hayashi, M., Okuyama, T., Aoki, H. \& Mizuno, H. A 20k-spin Ising chip to solve combinatorial optimization problems with CMOS annealing. {\it IEEE J. Solid-State Circuits} {\bf 51,} 303 (2016).
\bibitem{Aramon2019}Aramon, M., Rosenberg, G., Valiante, E., Miyazawa, T., Tamura H. \& Katzgraber, H. G. Physics-inspired optimization for quadratic unconstrained problems using a digital annealer. {\it Front. Phys.} {\bf 7,} 48 (2019).
\bibitem{Goto2019}Goto, H., Tatsumura, K. \& Dixon, A. R. Combinatorial optimization by simulating adiabatic bifurcations in nonlinear Hamiltonian systems. {\it Sci. Adv.} {\bf 5,} eaav2372 (2019).
\bibitem{Tatsumura2019}Tatsumura, K., Dixon, A. R. \& Goto, H. in FPGA-based simulated bifurcation machine. {\it 2019 29th International Conference on Field Programmable Logic and Applications (FPL)}, 59 (IEEE, New York, 2019).
\bibitem{Zou2020}Zou, Y. \& Lin, M. in Massively simulating adiabatic bifurcations with FPGA to solve combinatorial optimization. {\it Proceedings of the 2020 ACM/SIGDA International Symposium on Field-Programmable Gate Arrays (FPGA ’20)}, 65 (ACM, New York, 2020).
\bibitem{Goto2016}Goto, H. Bifurcation-based adiabatic quantum computation with a nonlinear oscillator network. {\it Sci. Rep.} {\bf 6,} 21686 (2016).
\bibitem{Goto2019a}Goto, H. Quantum computation based on quantum adiabatic bifurcations of Kerr-nonlinear parametric oscillators. {\it J. Phys. Soc. Jpn.} {\bf 88,} 061015 (2019).
\bibitem{Dykman2012}Dykman, M. {\it Fluctuating Nonlinear Oscillators: From Nanomechanics to Quantum Superconducting Circuits} (Oxford Univ. Press, Oxford, 2012).
\bibitem{Goto2018a}Goto, H., Lin, Z. \& Nakamura, Y. Boltzmann sampling from the Ising model using quantum heating of coupled nonlinear oscillators. {\it Sci. Rep.} {\bf 8,} 7154 (2018).
\bibitem{Amin2015}Amin, M. H. Searching for quantum speedup in quasistatic quantum annealers. {\it Phys. Rev. A} {\bf 92,} 052323 (2015). 
\bibitem{Goto2016a}Goto, H. Universal quantum computation with a nonlinear oscillator network. {\it Phys. Rev. A} {\bf 93,} 050301(R) (2016).
\bibitem{Puri2017a}Puri, S., Boutin, S. \& Blais, A. Engineering the quantum states of light in a Kerr-nonlinear resonator by two-photon driving. {\it npj Quant. Inform.} {\bf 3,} 18 (2017).
\bibitem{Puri2019a}Puri, S., St-Jean, L., Gross, J. A., Grimm, A., Frattini, N. E., Iyer, P. S., Krishna, A., Touzard, S., Jiang, L., Blais, A., Flammia, S. T. \& Girvin, S. M. Bias-preserving gates with stabilized cat qubits. Preprint at arXiv:1905.00450 (2019).
\bibitem{Dykman2018}Dykman, M. I., Bruder, C., L\"{o}rch, N. \& Zhang, Y. Interaction-induced time-symmetry breaking in driven quantum oscillators. {\it Phys. Rev. B} {\bf 98,} 195444 (2018).
\bibitem{Rota2019}Rota, R., Minganti, F., Ciuti, C. \& Savona, V. Quantum critical regime in a quadratically driven nonlinear photonic lattice. {\it Phys. Rev. Lett.} {\bf 122,} 110405 (2019).
\bibitem{Goto2019b}Goto, H., Lin, Z., Yamamoto, T. \& Nakamura, Y. On-demand generation of traveling cat states using a parametric oscillator. {\it Phys. Rev. A} {\bf 99,} 023838 (2019).
\bibitem{Yamamoto2008}Yamamoto, T., Inomata, K., Watanabe, M., Matsuba, K., Miyazaki, T., Oliver, W. D., Nakamura, Y. \& Tsai, J. S. Flux-driven Josephson parametric amplifier. {\it Appl. Phys. Lett.} {\bf 93,} 042510 (2008).
\bibitem{Lin2014}Lin, Z. R., Inomata, K., Koshino, K., Oliver, W. D., Nakamura, Y., Tsai, J. S. \& Yamamoto, T. Josephson parametric phase-locked oscillator and its application to dispersive readout of superconducting qubits. {\it Nat. Commun.} {\bf 5,} 4480 (2014).
\bibitem{Wang2019}Wang, Z., Pechal, M., Wollack, E. A., Arrangoiz-Arriola, P., Gao, M., Lee, N. R. \& Safavi-Naeini, A. H. Quantum dynamics of a few-photon parametric oscillator. {\it Phys. Rev. X} {\bf 9,} 021049 (2019).
\bibitem{Grimm2019}Grimm, A., Frattini, N. E., Puri, S., Mundhada, S. O., Touzard, S., Mirrahimi, M., Girvin, S. M., Shankar, S. \& Devoret, M. H. The Kerr-cat qubit: Stabilization, readout, and gates. Preprint at arXiv:1907.12131 (2019).
\bibitem{Nigg2017}Nigg, S. E., L\"{o}rch, N. \& Tiwari, R. P. Robust quantum optimizer with full connectivity. {\it Sci. Adv.} {\bf 3,} e1602273 (2017).
\bibitem{Puri2017}Puri, S., Andersen, C. K., Grimsmo, A. L. \& Blais, A. Quantum annealing with all-to-all connected nonlinear oscillators. {\it Nat. Commun.} {\bf 8,} 15785 (2017).
\bibitem{Zhao2018}Zhao, P., Jin, Z., Xu, P., Tan, X., Yu, H. \& Yu, Y. Two-photon driven Kerr resonator for quantum annealing with three-dimensional circuit QED. {\it Phys. Rev. Appl.} {\bf 10,} 024019 (2018).
\bibitem{Lechner2015}Lechner, W., Hauke, P. \& Zoller, P. A quantum annealing architecture with all-to-all connectivity from local interactions. {\it Sci. Adv.} {\bf 1,} e1500838 (2015).
\bibitem{Rocchetto2016}Rocchetto, A., Benjamin, S. C. \& Li, Y. Stabilizers as a design tool for new forms of the Lechner-Hauke-Zoller annealer. {\it Sci. Adv.} {\bf 2,} e1601246 (2016).
\bibitem{Pastawski2016}Pastawski, F. \& Preskill, J. Error correction for encoded quantum annealing. {\it Phys. Rev. A} {\bf 93,} 052325 (2016).
\bibitem{Albash2016a}Albash, T., Vinci, W. \& Lidar, D. A. Simulated-quantum-annealing comparison between all-to-all connectivity schemes. {\it Phys. Rev. A} {\bf 94,} 022327 (2016).
\bibitem{Hartmann2019}Hartmann, A. \& Lechner, W. Quantum phase transition with inhomogeneous driving in the Lechner-Hauke-Zoller model. {\it Phys. Rev. A} {\bf 100,} 032110(2019).
\bibitem{Susa2020}Susa, Y. \& Nishimori, H. Performance enhancement of quantum annealing under the Lechner-Hauke-Zoller scheme by non-linear driving of the constraint term. {\it J. Phys. Soc. Jpn.} {\bf 89,} 044006 (2020).
\bibitem{Chancellor2017}Chancellor, N., Zohren, S. \& Warburton, P. A. Circuit design for multi-body interactions in superconducting quantum annealing systems with applications to a scalable architecture. {\it npj Quant. Inform.} {\bf 3,} 21 (2017).
\bibitem{Leib2016}Leib, M., Zoller, P. \& Lechner, W. A transmon quantum annealer: decomposing many-body Ising constraints into pair interactions. {\it Quantum Sci. Technol.} {\bf 1,} 015008 (2016).
\bibitem{Note1}In numerical simulations of four KPOs in Ref.~\cite{Puri2017}, three KPOs are evolved in time via bifurcations, while another one is fixed in a coherent state.
\bibitem{Santoro2002}Santoro, G. E., Marto\v{n}\'{a}k, R., Tosatti, E. \& Car, R. Theory of quantum annealing of an Ising spin glass. {\it Science} {\bf 295,} 2427 (2002).
\bibitem{Leonhardt1997}Leonhardt, U. {\it Measuring the Quantum State of Light} (Cambridge Univ. Press, Cambridge, 1997).
\end{thebibliography}

\end{document}